\documentclass[11pt]{article}
\usepackage[a4paper]{geometry}
\usepackage{amssymb}
\usepackage{xy}
\usepackage{color} 
\xyoption{all}
\usepackage{mathrsfs}

\makeatletter

\def\qed{\ifvmode\removelastskip\fi
{\unskip\nobreak\hfil\penalty50\hbox{}\nobreak\hfil \hbox{\vrule
height1.2ex width1.2ex}\parfillskip=0pt \finalhyphendemerits=0
\par\medskip}}

\def\derpar#1#2{\mathchoice%
{{\partial#1\over\partial#2}}{{\partial#1/\partial#2}}%
{{\partial#1\over\partial#2}}{{\partial#1/\partial#2}}}

\makeatother

\def\ben{\begin{enumerate}}
\def\een{\end{enumerate}}
\def\beq{\begin{equation}}
\def\eeq{\end{equation}}
\def\bit{\begin{itemize}}
\def\eit{\end{itemize}}
\def\bea{\begin{eqnarray}}
\def\eea{\end{eqnarray}}
\def\beann{\begin{eqnarray*}}
\def\eeann{\end{eqnarray*}}

\newtheorem{teor}{Theorem}
\newtheorem{prop}{Proposition}
\newtheorem{lemma}{Lemma}
\newtheorem{cor}{Corollary}

\def\proof{\noindent\emph{Proof}\quad}

\let\ds=\displaystyle

\def\pr{{\rm pr}}
\def\R{\mathbb{R}}
\def\S{\mathbb{S}}

\def\cc{{\circ}}
\def\comp{\mathbin{\scriptstyle\circ}}

\def\tanvec#1{\derpar{}{#1}}

\def\Ker{\mathop{\rm Ker}\nolimits}

\def\Cinfty{{\rm C}^\infty}

\def\relat#1{\mathrel{\mathop{\sim}\limits_{#1}}}

\def\Tan{\mathrm{T}}
\def\Ver{\mathrm{V}}

\def\FD{\mathscr{F}}
\def\Lie{\mathscr{L}}
\def\d{\mathrm{d}}

\def\dfrac#1#2={{\displaystyle\frac{#1}{#2}}}

\title{\sffamily
Structural aspects of Hamilton--Jacobi theory
}

\author{\sffamily
Jos\'e~F. Cari\~nena$^a$, 
Xavier Gr\`acia$^b$, 
Giuseppe Marmo$^c$, 
\\
Eduardo Mart\'{\i}nez$^d$, 
Miguel~C. Mu\~noz--Lecanda$^b$, 
and 
Narciso Rom\'an--Roy$^b$
\\[2mm]
\normalsize
$^a$
Dept.\ Theoretical Physics and 
IUMA, Univ.\ Zaragoza
\\
\normalsize
$^b$
Dept.\ Mathematics, Univ.\ Polit\`ecnica de Catalunya, Barcelona
\\
\normalsize
$^c$
Dept.\ Physics, Univ.\ Federico~II di~Napoli
and
INFN, sezione di~Napoli
\\
\normalsize
$^d$
Dept.\ Applied Mathematics and 
IUMA, Univ.\ Zaragoza, 
}

\date{[\sl Int. J. Geom. Meth. Mod. Phys.
\bf 13\rm (2) (2016) 1650017 (29 pages).]}

\begin{document}

\abovedisplayskip=8pt plus 3pt minus 1pt
\belowdisplayskip=\abovedisplayskip
\belowdisplayshortskip=4pt plus 3pt minus 1pt

\thispagestyle{empty}%
{\sffamily\maketitle}

\begin{abstract}

\sffamily
\parindent 0pt
\noindent
In our previous papers
\cite{HJ,HJnh}
we showed that the Hamilton--Jacobi problem
can be regarded as a way to describe a given dynamics on a
phase space manifold
in terms of a family of dynamics on a lower-dimensional manifold.
We also showed how constants of the motion
help to solve the Hamilton--Jacobi equation.
Here we want to delve into this interpretation
by considering the most general case:
a dynamical system on a manifold
that is described in terms of a family of dynamics
(`slicing vector fields')
on lower-dimensional manifolds.
We identify the relevant geometric structures
that lead from this decomposition of the dynamics
to the classical Hamilton--Jacobi theory,
by considering special cases like fibred manifolds
and Hamiltonian dynamics,
in the symplectic framework and the Poisson one.
We also show how a set of functions on a tangent bundle
can determine a second-order dynamics for which they are
constants of the motion.

\bigskip
\slshape


MSC\,2010: 70H20, 70G45

\end{abstract}

\section{Introduction}

Hamilton--Jacobi theory originated with Hamilton
to deal with what nowadays is called Hamiltonian optics,
i.e.\
to describe the ray propagation of light,
and with Jacobi who was interested
in devising a procedure to integrate equations of motions
when they are given in canonical form.
In Jacobi's own words:
``After we have reduced the problems of mechanics to
the integration of a nonlinear first order
partial differential equation, we must concern
ourselves with the integration of the same,
i.e., with the search for a complete solution''
\cite[p.\,183]{Jac}.
Hadamard \cite{Had08, Had23}
and Volterra \cite{Vol59}
derived the Hamilton--Jacobi equations
by considering the short-wave limit of wave equations.
It was this association which paved the way for de Broglie
to introduce the relation
$$
p\,\d x-H\, \d t = \hbar\,(k\,\d x-\omega\,\d t)
$$
relating wave concepts with particle concepts \cite{AMo7}.
Using this analogy,
Schr\"{o}dinger proposed the evolutionary equation for wave mechanics,
opening the route to a formalism
able to describe physical phenomena at atomic scale.
(A geometrical description of the quantum--to--classical transition
on space--time was elaborated by Synge
\cite{S54}.)

Concerning the role of solutions to the Hamilton--Jacobi equation,
providing a family of solutions for Hamilton's equations,
Dirac wrote
\cite{Dirac}:
``The family does not have any importance
from the point of view of Newtonian mechanics;
but it is a family which corresponds to one state of motion
in the quantum theory,
so presumably the family has some deep significance in nature,
not yet properly understood''.
These general comments are aimed at contextualizing
the role of the Hamilton--Jacobi theory in theoretical physics.
To enter the
\emph{raison d'\^etre} of the present paper,
let us recall how Hamilton--Jacobi theory is usually dealt with
in textbooks and works on analytical mechanics
\cite{Ar,Go51,MMMcq,Sa71,JoseSaletan,Wi37}.

Hamilton--Jacobi theory is usually considered
when dealing with canonical transformations
to define them by means of generating functions.
Specifically, by using canonical coordinates,
say $(p,q;t)$ and $(\bar p,\bar q;t)$,
one looks for a function
$S \colon Q \times Q \times \R \to \R$
such that
$$
p\,\d q-H\,\d t
=
\bar p\,\d \bar q-K\,\d t+\d S(q,\bar q;t)
\,,
$$
with $H$ and $K$ Hamiltonian functions on phase space.
The associated transformation is defined by means of
the implicit equations
$$
p = \frac{\partial S}{\partial q}
\,,\quad
\bar p = -\frac{\partial S}{\partial \bar q}
\,,\quad
K-H = \frac{\partial S}{\partial t}
\,,
$$
and this canonical transformation,
if it exists, 
converts the Hamiltonian system described by~$H$
into the one described by~$K$.
By further requiring that $K$ is a constant
or that it is a function depending only on~$\bar p$,
one relates the original system
to another one which is completely integrable,
and therefore integrable by quadratures.

The short-wave limit point of view 
starts from a second order partial differential equation
of hyperbolic type and derives what is known
as the \emph{eikonal equation}
$$
\left(\frac{\partial S}{\partial x}\right)^2+
\left(\frac{\partial S}{\partial y}\right)^2+
\left(\frac{\partial S}{\partial z}\right)^2
=n^2(x,y,z)
$$
with $n$ denoting the refractive index
\cite[p.\,119]{BW}
\cite[p.\,108]{EMS}.
The function $S$ is usually called the
\emph{eikonal function}
or the characteristic function.
As a matter of fact,
Hamilton introduced two functions,
$S(t,x,y,z)$, called the \emph{principal function},
and putting $W(x,y,z)-t\, E=S(t,x,y,z)$,
$W$ was called the \emph{characteristic function} \cite {LV14}.

From the point of view of Jacobi
\cite{Jac},
the integration of Hamilton's equations is achieved
by solving first the first-order differential equation
on configuration space
$$
\frac{\d  q^j}{\d  t}
=
\left. \frac{\partial H}{\partial p_j} \right|%
_{p_j=\frac{\partial S}{\partial q^j}}
\,;
$$
then, setting
$$
p_j=\frac{\partial S}{\partial q^j}(t,q^j(t))
\,,
$$
one finds a full solution of Hamilton's equations
with initial condition $(q^j(0), p_j(0), t=0)$.
Thus, from this point of view,
the Hamilton--Jacobi equation is instrumental
to define a
\emph{family of first--order differential equations}
on 
\emph{configuration space}
whose solutions will eventually produce solutions
for the Hamilton  equations on phase space.
In the first-order differential equation
(\emph{fode}, in the sequel)
$$
\frac{\d  q^j}{\d  t}
=
\left. \frac{\partial H}{\partial p_j} \right|%
_{p_j=\frac{\partial S}{\partial q^j}}
$$
one changes the values
of the arbitrary constants appearing
in a complete integral function~$S$
and obtains a family of \emph{fode}s.
The solutions of each one of these equations
are the solutions alluded to by Dirac
and correspond to a given~$S$,
related to the phase of the wave function in quantum mechanics.

Therefore a complete solution to the Hamilton--Jacobi equation
gives rise to a family of first-order differential equations
on the configuration space,
say~$Q$,
which are sufficient to recover
all the solutions to Hamilton's equations on $\Tan^*Q$.
From the geometrical point of view,
a complete solution amounts to
an invariant foliation of $\Tan^*Q$,
with leaves diffeomorphic to~$Q$
and transverse to fibres of the cotangent bundle projection.
A family of first--order differential equations is obtained
by restricting the Hamiltonian vector field
to each leaf of the invariant foliation.

From all that we have said about Hamilton--Jacobi theory
it is clear that we may identify
two main aspects in the Hamilton--Jacobi theory.
The first one is to solve a \emph{fode} in a manifold~$P$
(usually $\Tan^*Q$)
by solving an associated family of
\emph{fode}'s on a lower dimensional manifold~$Q$;
when all the solutions may be found in this manner, the family is
said to be \emph{complete}.
The second one consists of finding this complete family by solving
an associated PDE for a single function~$S$,
this would be the analog of the \emph{eikonal equation}.

\smallskip

To analyse these problems 
we introduce a general scheme by means of
a vector field $Z$ on a manifold~$P$,
along with a fibration $P\to M$.
We consider all the integral curves of $Z$ on $P$ 
and project them onto~$M$.
Having all these curves on $M$, we would like to `group' them
into coherent sets of integral curves for vector fields on~$M$.
In other terms, we would like to put together
all those integral curves of $Z$ which may be obtained as
integral curves of a certain vector field $X$ on~$M$.
If all integral curves of $Z$ may be grouped into families
such that each family, after projection,
arises as integral curves of a vector field $X$ on~$M$,
we say that the family of vector fields~$X$ 
is a
\emph{complete slicing} 
of the dynamics~$Z$,
or that it is a complete solution to the
\emph{generalized Hamilton--Jacobi problem}. 
This paper deals mostly with the first aspect, 
i.e., to solve a differential equation on~$P$ by means of
a familly of differential equations on~$Q$.

A similar problem, i.e.\
going from trajectories to vector fields on~$M$,
was discussed in 
\cite[chapter 6]{MSSV}.
It is shown there that, in this generality,
by no means the problem will have solutions.
Thus, the existence of a family of vector fields on~$M$
sufficient to reproduce all integral curves of $Z$ on~$P$
will put quite strong conditions on~$Z$.
We have already remarked that
for Hamiltonian systems on $P=\Tan^*Q$
the existence of the family
would require $Z$ to be a completely integrable system.
Of course a kind of inverse problem could be posed:
given a family of vector fields on~$M$,
is it possible to find a vector field $Z$ on~$P$
such that it would be possible to represent
the whole family of integral curves of the various vector fields on~$M$
as projections of integral curves of the vector field $Z$ on~$P$?
Let us stress that these problems would arise
in particular physical problems
like motion of particles with internal structure and 
in general in problems with 
restricted allowed Cauchy data, 
for instance, gauge terms. 
It would also occur in quantum mechanics
when we consider a composite system and
we would like to describe it
in terms of the evolution of subsystems
(entanglement would be an obstruction to the solution of
the posed inverse problem).
A single case where the inverse problem has a nice solution
is provided by a second order vector field on $\Tan Q$,
completely determined by a suitable family of
functionally independent constants of the motion,
as we show at the end of the paper.

To pin-point the geometrical contents
of the standard Hamilton--Jacobi equation,
first we shall consider the usual Hamilton--Jacobi
theory from a more geometric point of view.
In the usual approach
$P=\Tan^*Q$,
and $\pi \colon P\to Q$ is the usual cotangent bundle projection.
The dynamical vector field $\Gamma=Z$
solves the equation $\mathrm{i}_{\Gamma}\omega=\d H$,
where
$\omega$ is the canonical symplectic structure in $\Tan^*Q$ and
$H$ is the Hamiltonian function.
By using the symplectic potential for $\omega$,
say $\omega=-\d \theta_0$,
we define a vector field $\Delta$,
$\mathrm{i}_{\Delta}\omega=
\theta_0$,
which represents the
linear structure along the fibers,
and the Hamilton--Jacobi equation for $S$ becomes
$$
(\d S)^*\theta_0=\d S
\,,\quad
(\d S)^*H=E
\,,
$$
where $E$ is a `parameter'.
When $S$ is a complete integral,
we have that
$\d S \colon Q \times N \to \Tan^*Q$
is a diffeomorphism for `most initial conditions'
for~$\Gamma$.
It provides a $\dim Q$-foliation of $\Tan^*Q$
(or some open dense submanifold of it)
transversal to the fibers.
The vector field $\Gamma$, restricted to each leaf,
being tangent to it,
defines a vector field
which projects onto a vector field $X$ defined on $Q$.
There would  be a vector field $X$ for each leaf.
In this manner the invariant foliation
 defines a family of first-order differential equations on $Q$,
each one of them being the projection
of the restriction of $\Gamma$ to the invariant leaf.
This means that $\Gamma$ may be replaced by
the family of vector fields that we obtain
by restricting $\Gamma$ to a family of leaves transversal to the fibres.
Thus the issue becomes
how to find an invariant foliation transversal to fibres.

\smallskip
These and other intrinsic considerations about
the Hamilton--Jacobi equation can be found in
\cite{AM,LM,MMM}.
In addition, in
\cite{HJ} a general geometric framework for the Hamilton--Jacobi theory
was presented
and the Hamilton--Jacobi equation in the
Lagrangian and in the Hamiltonian formalisms was formulated
for autonomous and non-autonomous mechanics,
recovering the usual Hamilton--Jacobi equation as a special case
in this generalized framework.
The relationship between the Hamilton--Jacobi equation
and some geometric structures of mechanics were analyzed also in
\cite{BLM-12,HJteam-K}.
A similar generalization of the Hamilton--Jacobi formalism
was outlined in
\cite{KV-1993}.
Later on, these geometric frameworks were used to develop the
Hamilton--Jacobi theory in many different situations.
Thus, in
\cite{BFS-14,HJnh,leones1,leones2,blo,OFB-11}
this is done for holonomic and non-holonomic mechanical systems,
in \cite{Gomis1,LOS-12,LMV-12,LMV-13}
the theory is extended for singular systems,
in \cite{BMMP-10,LS-12} and
\cite{LMV-14}
for geometric mechanics on Lie algebroids and almost-Poisson manifolds
respectively,
in \cite{BLMMM-12,Wang1,Wang2}
for control theory,
in \cite{CLMV-14,LMMSV-12,LPRV-15,DeLeon_Vilarino,Vi-12}
for different formulations of classical field theories
(and in \cite{Vi-11} for partial differential equations in general),
and in
\cite{CLPR1,CLPR2,art:Constantelos84,Vi-10}
for higher order dynamical systems and higher-order field theories.
Finally, the geometric discretization of the Hamilton--Jacobi equation
is also considered in
\cite{BDM-12,OBL-11}.

In particular, in our previous papers
\cite{HJ,HJnh}
we saw that the Hamilton--Jacobi problem
can be regarded as a way
to describe a given dynamics on a phase space manifold
in terms of a family of dynamics on a lower-dimensional manifold.
Moreover, we saw that the existence of many constants of the motion
for the given dynamics helps to solve the Hamilton--Jacobi problem.
The aim of this paper is 
to look more deeply into
this interpretation
by considering the most general case and
identifying what are the relevant geometric structures.

We should remark that our framework allows to handle
dynamical vector fields
which cannot be handled with classical approaches to 
Hamilton--Jacobi equation.
For instance, 
suppose we have a completely integrable Hamiltonian system
given by a Hamiltonian vector field $Z_H$;
its Hamilton--Jacobi equation has a complete solution,
and therefore we have a complete slicing of the dynamics.
Then consider a new dynamics given by
$Z' = f\,Z_H$,
where $f$ is a generic function
---this leads to a reparametrization of the integral curves.
Our procedure allows to construct a complete slicing for~$Z'$,
although $Z'$ may not be Hamiltonian.
(An instance where this reparametrization may be required 
is when $Z_H$ is not a complete vector field.)

\medskip

The paper is organized as follows:
In section~2 we present the general concepts and results needed to state
a more general framework for the Hamilton--Jacobi problem.
The study of constants of the motion and complete solutions and
their relationship for this general setting is done in section~3,
by introducing the concept of \emph{slicing vector fields}
and complete slicings.
Section~4 is devoted to discuss some particular situations deriving from
this general framework, such as Hamiltonian systems defined on
symplectic and Poisson manifolds.
The slicing problem is discussed again in section~5 in the case
where the dynamical system, either general or Hamiltonian,
is defined on a generic fibered manifold.
Finally, in section~6
we show how our previous results in
\cite{HJ} are recovered form here,
and we also study how the knowledge of enough constants of the motion
determines a second-order dynamics.
Along the work,
different examples are also introduced to illustrate our results.
All the manifolds and maps are assumed to be~$\Cinfty$.

\section{Dynamical systems, invariant submanifolds and constants of the motion}

\subsection*{Dynamical systems}

A dynamical system is a pair $(P,Z)$
given by a manifold $P$ and a vector field $Z$ on~$P$.
This defines a (first-order, autonomous) differential equation
on~$P$,
$\gamma' = Z \comp \gamma$,
for a path $\gamma \colon I \to P$.
This dynamics may possess several features.
For the purposes of this work we are especially interested in
invariant submanifolds and constants of the motion.

A submanifold $M \subset P$ is said to be invariant by~$Z$
when the flow of $Z$ leaves $M$ locally invariant,
or, in other words,
when every integral curve of $Z$ meeting $M$ is
contained in $M$ at least for some time
(if $M$ is not closed then this integral curve may
eventually leave it).
These conditions are equivalent to saying that
$Z$ is tangent to~$M$.
The preceding definition is applicable to regular submanifolds
but also to immersed submanifolds.

A particular instance of invariant submanifolds
is provided by constants of the motion.
In its most elementary form
a constant of the motion for $Z$ is a function
$f \colon P \to \R$
such that,
along every integral curve $\gamma$ of~$Z$,
the function $f \comp \gamma$ is constant.
This is equivalent to saying that
the Lie derivative of $f$ with respect to~$Z$ is zero,
$\Lie_Z f = 0$.
In the same way one can consider
a vector-valued constant of the motion
$F \colon P \to \R^n$,
whose components are scalar constants of the motion,
or, more generally,
a manifold-valued map
$F \colon P \to N$
such that for every integral curve $\gamma$
the map $F \comp \gamma$ is constant.
If $c \in N$,
then the closed subset $F^{-1}(c) \subset P$
is clearly invariant by~$Z$.
So, those of the sets $F^{-1}(c)$ that are not empty
constitute a partition of~$P$.
In some cases we can ensure that they are also submanifolds,
for instance when $F$ is a submersion.
In this case, constants of the motion provide a whole family
of invariant submanifolds.

Of course, not all invariant submanifolds are levels sets of
constants of the motion.
A very simple example is given by the planar system
$\dot x = -y + x(1\!-\!x^2\!-\!y^2)$,
$\dot y =  x + y(1\!-\!x^2\!-\!y^2)$,
that reads in polar coordinates
$\dot r = r(1\!-\!r^2)$,
$\dot\phi = 1$;
it has an equilibrium point (the origin),
a limit cycle ($r=1$),
and no nontrivial \emph{global} constants of the motion.
More interesting examples are provided by Li\'enard's equation
and the particular case given by
van der Pol's equation.
For instance, the system
$\dot x = -y+x \,\sin(x^2+y^2)$,
$\dot y = x+y \,\sin(x^2+y^2)$,
has a countable number of limit cycles.

\subsection*{A general framework for the Hamilton--Jacobi theory:
slicing vector fields}

One of the distinctive facts of the Hamilton--Jacobi equation
is that it allows to describe the dynamics given by the
Hamilton equation on the cotangent bundle
in terms of a family of first-order dynamics on the configuration
space
(as for instance in
\cite[theorem 5.2.4]{AM}).
According to this general principle,
to describe the dynamics $Z$ on~$P$
in terms of other dynamics on lower-dimensional manifolds,
we consider another
manifold~$M$,
a vector field $X$ on~$M$,
and a map $\alpha \colon M \to P$.
The following diagram captures the situation:
$$
\xymatrix{
*++{\Tan M}  \ar[r]^{\Tan \alpha}  \ar[d]_{} &
*++{\Tan P}  \ar[d]_{}
\\
*++{M} \ar[r]^{\alpha} \ar@/^3mm/[u]^{X} &
*++{P} \ar@/_3mm/[u]_{Z}
}
$$
What can be said about the relation between $X$, $\alpha$ and~$Z$?
The following results are well-known:
\begin{prop}
Given the preceding data,
the following properties are equivalent:
\begin{enumerate}
\itemsep 0pt plus 1pt
\item
For every integral curve $\xi$ of~$X$,
$\zeta = \alpha \comp \xi$ is an integral curve of~$Z$.
\item
$X$ and $Z$ are $\alpha$-related
($X \relat{\alpha} Z$),
that is to say,
\beq
\label{slicing}
\Tan \alpha \comp X = Z \comp \alpha
\,,
\eeq
\end{enumerate}
Suppose moreover that $\alpha$ is an \emph{injective immersion},
thus inducing a diffeomorphism
$\alpha_\cc \colon M \to \alpha(M)$
of $M$ with an immersed submanifold
$\alpha(M) \subset P$.
Then the preceding properties are also equivalent to
\begin{enumerate}
\itemsep 0pt plus 1pt
\setcounter{enumi}{2}
\item
$Z$ is tangent to $\alpha(M)$,
and, if $Z_\cc$
is the restriction of $Z$ to $\alpha(M)$,
$X$ is given by the pullback
$X = \alpha_\cc^*(Z_\cc)$.
\end{enumerate}
In this case,
the map
$\xi \mapsto \alpha \comp \xi$
is a bijection between integral curves of~$X$
and integral curves of~$Z$ passing through $\alpha(M)$.
\qed
\end{prop}

When these conditions hold,
we can regard $X$ as a `partial dynamics',
or a `slice' of the dynamics given by~$Z$.
Eventually,
if we knew enough of these slices,
we could recover the whole dynamics of~$Z$.

\paragraph{Definition}
{\itshape
Given a dynamical system $(P,Z)$,
we will call a
\emph{slicing} of it
a triple
$(M,\alpha,X)$
satisfying the
\emph{slicing equation}~(\ref{slicing}).}
\medskip

When $\alpha$ is an immersion
the vector field $X$, if it exists,
is uniquely determined by $\alpha$ and $Z$;
so, in this case, we can speak of $(M,\alpha)$
being a solution of the slicing equation for $(P,Z)$.
This hypothesis 
will hold in many applications,
in particular for the sections $\alpha$ of a bundle $P \to M$
(as a matter of fact, they are embeddings).

As we will see later on in this paper,
equation~(\ref{slicing}) may be thought of as a
generalisation of the
Hamilton--Jacobi equation.
One of our main purposes is
to identify the precise conditions
that take us from the slicing equation to
the Hamilton--Jacobi equation.

\paragraph{Coordinate expression}
Let us express equation~(\ref{slicing}) in coordinates.
Consider coordinates
$(x^i)$ in~$M$,
$(z^k)$ in~$P$,
and use them to express
the map
$\alpha(x) = (a^k(x))$
and
the vector fields
$X = X^i \,\tanvec{x^i}$,
and
$Z = Z^k \,\tanvec{z^k}$.
Then the difference
$
\Tan \alpha \comp X - Z \comp \alpha
$
reads
$$
(x^i) \mapsto
\left(
a^k(x) ,
\derpar{a^k}{x^i} \, X^i -
Z^k(\alpha(x))
\right)
\,,
$$
and so
$(M,\alpha,X)$ is a solution of the slicing equation iff
$\ds
\derpar{a^k}{x^i} \, X^i(x) =
Z^k (\alpha(x))
$.

\subsection*{Gauge freedom of the solutions}

The notion of a slicing of~$Z$ has a certain
`gauge freedom',
in the sense that with a given solution $(M,\alpha,X)$
there exist many associated solutions that are
\emph{equivalent} to it:
if $\varphi \colon M' \to M$ is a diffeomorphism
then
$(M',\alpha \comp \varphi,\varphi^*(X))$
is also a solution of the slicing equation.
There are two situations where this freedom can be easily removed.

One, to be studied later on,
occurs when $P$ is assumed to be fibred over a manifold
and one only deals with maps $\alpha$
that are sections of this projection.

The other one is provided by invariant submanifolds of~$P$.
Indeed, this is an immediate consequence of proposition~1:
\begin{cor}
Let
$P_\cc \subset P$
be a regular submanifold.
The canonical inclusion
$j \colon P_\cc \hookrightarrow P$
is a solution of the slicing equation
iff
$Z$ is tangent to~$P_\cc$.
\\
Every other solution given by an embedding $\alpha$
with $\alpha(M) = P_\cc$
is equivalent to it.
\qed
\end{cor}

\section{Constants of the motion and complete solutions}

\subsection*{Constants of the motion}

We still deal with our dynamical system $(P,Z)$.
A (generalized)
\emph{constant of the motion}
of it is a map
$F \colon P \to N$ into another manifold $N$
satisfying the following property:
for any integral curve
$\zeta \colon I \to P$ of~$Z$,
$F \comp \zeta$ is constant.

\paragraph{Example}
We consider the isotropic harmonic oscillator 
with two degrees of freedom 
(with phase space $\R^4$),
$$
\begin{array}{rcl} \dot x&=& -y\\
\dot y&=&x.
\end{array}
$$
All its integral curves are a foliation of
$\R^4 -\{0\} \cong \S^3 \times \R^+$ onto
$\R^3 -\{0\} \cong \S^2 \times \R^+$
and the projection
$\R^4 -\{0\} \to \R^3 -\{0\}$ 
(Kustaanheimo--Stiefel map),
or $\S^3 \to \S^2$,
is a constant of the motion.

\medskip

We have several characterisations of this property:
\begin{prop}
The following properties are equivalent:
\begin{enumerate}
\itemsep 0pt plus 1pt
\item
$F$ is a (manifold valued) constant of the motion.
\item
Each integral curve $\eta$ of~$Z$ is contained in a level set
$F^{-1}(c)$ of~$F$.
\item
$Z$ is $F$-related with the zero vector field of~$N$:
($Z \relat{F} 0$).
\end{enumerate}
Suppose moreover that $F$ is a \emph{submersion}
(thus $\Ker \Tan F \subset \Tan P$
is an integrable tangent subbundle
whose associated foliation has as leaves the
level sets $F^{-1}(c)$,
which are closed submanifolds of~$P$).
Then the preceding properties are also equivalent to
\begin{enumerate}
\setcounter{enumi}{3}
\itemsep 0pt plus 1pt
\item
$Z$ takes its values in $\Ker \Tan F$.
\item
$Z$ is tangent to every level set $F^{-1}(c)$.
\qed
\end{enumerate}
\end{prop}
%
The following diagram summarizes the situation:
$$
\xymatrix{
&
*++{\Tan P}  \ar[r]^{\Tan F}  \ar[d]_{} &
*++{\Tan N}  \ar[d]_{}
\\
*++{I} \ar[r]^{\eta} &
*++{P} \ar[r]^{F} \ar@/^3mm/[u]^{Z} &
*++{N} \ar@/_3mm/[u]_{0}
}
$$

The tangency of~$Z$ to a certain submanifold
shows up in propositions 1 and~2.
This comparison suggests that
a constant of the motion
is related to a whole family of solutions of
the slicing equation,
as we are going to show.

\subsection*{Complete solutions}

A single solution
$\alpha \colon M \to P$, $X \colon M \to \Tan M$,
of the slicing equation
allows to describe the integral curves of $Z$
contained in $\alpha(M) \subset P$.
To describe \emph{all} of its integral curves
we need a \emph{complete} solution.
This can be defined as a family of solutions indexed by some
parameter space~$N$.

\paragraph{Definition}
{\itshape
Given a dynamical system $(P,Z)$, a
\emph{complete slicing}
of it is given by
\begin{itemize}
\itemsep 0pt
\item
a map
$
\overline\alpha \colon M \times N \to P
$
and
\item
a vector field
$
\overline X \colon M \times N \to \Tan M
$
along the projection
$M \times N \to M$
\end{itemize}
(that is, smooth families of
maps
$\alpha_c \equiv \overline\alpha(\cdot,c) \colon M \to P$
and vector fields
$X_c \equiv \overline X(\cdot,c) \colon M \to \Tan M$,
both indexed by the points $c \in N$)
such that:
\begin{itemize}
\itemsep 0pt
\item
$\overline\alpha$ is surjective
(or at least its image is an open dense subset),
and
\item
for each $c \in N$,
the map
$\alpha_c 
\colon M \to P$
and the vector field
$X_c 
\colon M \to \Tan M$
constitute a slicing of~$Z$.
\end{itemize}%
}%
$$
\xymatrix{
*++{\Tan M \times N} \ar[r]^{\Tan_1 \overline\alpha} \ar[d]_{} &
*++{\Tan P} \ar[d]_{}
\\
*++{M \times N} \ar[r]^{\overline\alpha} \ar@/^3mm/[u]^{\overline X} &
*++{P} \ar@/_3mm/[u]_{Z}
}
$$
Since (almost) every $z \in P$ is the image by~$\overline\alpha$ of a point
$(x,c) \in M \times N$,
the integral curve of $Z$ through~$z$
can be described as the integral curve of $X_c$ through $x$
by means of the map $\alpha_c$.

When each $\alpha_c$ is an immersion
(for instance, when $\overline\alpha$ is a diffeomorphism)
the vector fields $X_c$ are determined by the $\alpha_c$,
so in this case we do not need to specify $\overline X$
to define the complete solution.

\paragraph{Example}
The simplest example of a solution of the slicing equation
for a vector field $Z$ is just given by its integral
curves $\alpha \colon I \to P$.
Indeed, consider the following diagram:
$$
\xymatrix{
*++{\Tan I}  \ar[r]^{\Tan \alpha}  \ar[d]_{} &
*++{\Tan P}  \ar[d]_{}
\\
*++{I} \ar[r]^{\alpha} \ar@/^3mm/[u]^{{\d \over \d t}}
\ar[ru]^{\alpha'}
&
*++{P} \ar@/_3mm/[u]_{Z}
}
$$
The commutativity of its upper triangle is
the definition of the velocity $\alpha'$,
whereas the commutativity of the lower one
is the assertion that $\alpha$ being an integral curve of~$Z$.
When this holds,
${\d \over \d t} \relat{\alpha} Z$,
which means that $\alpha$ is a solution of the
slicing equation for~$Z$.

Let $z \in P$ be a \emph{noncritical point} of~$Z$.
Then one can build a
\emph{local} complete slicing
around~$z$.
To this end, consider a hypersurface
$N \subset P$ containing $z$,
and such that $Z(z)$ is transversal to~$N$.
Then the restriction of the flow $F$ of $Z$
to a smaller product $I_\cc \times N_\cc$
gives a diffeomorphism
$F_\cc \colon I_\cc \times N_\cc \to P_\cc$
with an open neighbourhood
$P_\cc$ of~$z$,
such that
${\partial \over \partial t} \relat{F_\cc} Z$.
So, $F_\cc$ with ${\partial \over \partial t}$
is a complete slicing for~$Z$
restricted to $P_\cc$.
Indeed, this is the usual procedure to prove
the straightening theorem for vector fields.

\subsection*{Local existence of complete slicings}

The preceding example can be extended
to prove a general existence theorem for complete slicings.
Indeed, we are going to prove that,
under some regularity conditions,
\emph{any} given slicing
can be \emph{locally} embedded in a regular local complete slicing.

\begin{teor}
Let $(P,Z)$ be a dynamical system,
and $z_\cc \in P$ a \emph{noncritical} point of~$Z$.
Let $(M,\alpha,X)$ be a solution
of the slicing equation for~$Z$,
with $z_\cc = \alpha(x_\cc)$,
and such that
$\alpha$ is an immersion at~$x_\cc$.

There exist
an open neighbourhood $M_\cc$ of~$x_\cc$,
an open neighbourhood $N_\cc$ of~0 in $\R^n$
(where $n = \dim P - \dim M$),
and
a diffeomorphism
$\overline\alpha \colon M_\cc \times N_\cc \to P_\cc$
with an open neighbourhood $P_\cc$ of~$z_\cc$,
such that
\begin{itemize}
\itemsep 0pt plus 1pt
\item
$\overline\alpha$ is a complete slicing
for $Z|_{P_\cc}$,
and
\item
$\overline\alpha(\cdot,0) = \alpha|_{M_\cc}$.
\end{itemize}
\end{teor}

\proof
Since the result is a local one,
and every immersion is locally an embedding,
the gauge freedom of the solutions of the slicing equation
allows us to suppose that $M$ is a regular submanifold of~$P$
and that $\alpha$ is the inclusion.
The hypothesis is that $Z$ is tangent to~$M$.

The proof of the straightening theorem for vector fields
can be adapted
to construct coordinates
$(z_1,\ldots,z_m,\ldots,z_p)$ around~$z_\cc$
such that
$M$ is locally described by
$z_{m+1} = \ldots = z_p = 0$,
and that
$Z = \tanvec{z_1}$.
Then,
in a small product $M_\cc \times N_\cc$,
define
$\overline\alpha(x;s_1,\ldots,s_n)
=
(z_1(x),\ldots,z_m(x),s_1,\ldots,s_n)$,
where the right-hand side is expressed in terms of
these coordinates.
In a small neighbourhood of $(x_\cc,0)$
this is a diffeomorphism,
and for every $s \in N_\cc$
the vector field $Z$ is tangent to the submanifold
$\alpha_s(M_\cc)$.
Therefore $\overline{\alpha}$ is a complete slicing of~$Z$.
\qed


\subsection*{Relation between complete slicings,
constants of the motion and connections}

Now we are going to see that,
under some regularity hypotheses,
there is a close relationship between
complete slicings and constants of the motion.

\begin{teor}
\label{teor-complete-constant}
Let $(P,Z)$ be a dynamical system,
and
$
\overline\alpha \colon M \times N \to P
$
a diffeomorphism.
Then
$\overline\alpha$
is a complete slicing
for~$Z$
iff
$F = \pr_2 \comp \overline\alpha^{-1} \colon P \to N$
is a constant of the motion for $Z$.
$$
\xymatrix{
*++{P}  \ar[r]^{F}  &
*++{N}
\\
*++{M \times N} \ar[u]^{\overline\alpha\,}
\ar[ru]_{\mathrm{pr}_2}
}
$$
\end{teor}

\proof
If $\overline\alpha$ is a complete slicing,
for each $c \in N$,
$\overline\alpha$ restricts to a map
$M \times \{c\} \to \alpha_c(M)$
which is a diffeomorphism,
and all the integral curves of~$Z$ in $\alpha_c(M)$
correspond to a common value of~$c$.
This means the map
$F = \pr_2 \comp \overline\alpha^{-1} \colon P \to N$
is a constant of the motion.

Conversely,
from $F = \pr_2 \comp \overline\alpha^{-1}$
we have that, for every $c$,
$F(\overline\alpha(x,c)) = c$,
or $\alpha_c(M) \subset F^{-1}(c)$.
Both submanifolds have the same dimension,
and,
since $F$ is a constant of the motion,
$Z$ is tangent to
$F^{-1}(c)$;
therefore $Z$ is tangent to
$\alpha_c(M)$,
which proves that
the $\alpha_c$ are solutions to slicing equation for~$Z$.
\qed

This result shows that there is a \emph{bijection}
between complete slicings and constants of the motion,
but these are being assumed to satisfy a very strong
regularity condition,
which essentially requires that
all the level sets $F^{-1}(c)$
are diffeomorphic to a common manifold $M$,
in such a way that gluing the collection of diffeomorphisms
$M \to F^{-1}(c)$
yields a diffeomorphism
$\overline\alpha \colon M \times N \to P$.
Of course,
these conditions are very restrictive,
but in practice they may hold in a generic way.
We will see this in some examples.

\paragraph{Example}
Consider the manifold $\R^2$ with the radial vector field
$Z = z_1 \,\tanvec{z_1} + z_2 \,\tanvec{z_2}$,
whose integral curves are
the equilibrium at the origin and
the paths $\zeta(t) = e^t (a_1,a_2)$,
$(a_1,a_2) \neq (0,0)$,
running along the half-lines from the origin.

To illustrate the preceding theorem
we have to exclude the origin:
$P = \R^2-\{0\}$.
The map $F \colon P \to N = \S^1$ given by
$F(z) = z/\|z\|$
is clearly a constant of the motion for~$Z$.
Its level sets $F^{-1}(u)$ (for $u \in \S^1$)
are diffeomorphic to the real line~$M = \R$;
for instance, by
$\alpha_u \colon \R \to P$,
$\alpha_u(x) = e^x u$.
All together yield a diffeomorphism
$\overline\alpha \colon \R \times \S^1 \to \R^2-\{0\}$:
$\overline\alpha(x,u) = e^x u$.
This is a complete solution of the slicing equation for~$Z$.
The corresponding vector fields on $M$ are
$X_u = \tanvec{x}$.
\medskip

The relationship between slicings and constants of the motion
is lost when we do not consider complete slicings.
A solution of the slicing equation doesn't need
to preserve any given constant of the motion,
and the preservation of a constant of the motion
does not guarantee that
a map is a slicing of the dynamics.
The simplest way to show all this is by an example.

\paragraph{Example}
We consider the manifold
$P = \R^3$,
with coordinates $(x,y,z)$,
and the simple dynamics given by the vector field
$Z = \tanvec{x}$.
The function $F = z$ is obviously a constant of the motion
with values in $\R$.

The map $\alpha \colon \R^2 \to \R^3$
given by $\alpha(u,v) = (u,v,0)$,
satisfies
$F \comp \alpha = 0$, constant.
On the other hand,
$\bar \alpha(u,v) = (u,0,v)$
satisfies  $(F \comp \bar\alpha) (u,v) = v$,
not constant.
Both $\alpha$ and $\bar\alpha$
are solutions of the slicing equation for $(P,Z)$,
since $Z$ is tangent both to the planes
$\alpha(\R^2)$ and $\bar \alpha(\R^2)$.

Now consider
$\beta \colon \R \to \R^3$
given by
$\beta(v) = (0,v,0)$.
Obviously $F \comp \beta = 0$
but $\beta$ is not a solution of the slicing equation
since $Z$ is not tangent to the line $\beta(\R)$.

\subsection*{Invariant foliations}

The notion of complete solution is close to that of invariant
foliation.
Roughly speaking,
a foliation of $P$ consists in describing it
as the disjoint union of immersed submanifolds.
This defines an integrable tangent distribution on~$P$,
and conversely.
The leaves of the foliation
are solutions of the slicing equation for~$Z$
iff
$Z$ is tangent to the foliation
(or, in other words, if the foliation is invariant by the flow
of~$Z$).
Equivalently,
iff $Z$ is a section of the associated tangent distribution.

So, if $\overline\alpha \colon M \times N \to P$
is a complete slicing,
bijective,
and with every partial map $\alpha_c$ an immersion,
then the submanifolds $\alpha_c(M)$ are a foliation of~$P$
invariant by~$Z$.
However, not every invariant foliation
can be defined by a \emph{global} diffeomorphism in this way.

\paragraph{Example}
Consider the `irrational linear flow' on the
2--dimensional torus $\mathbb{T}^2$:
$\dot x = 1$, $\dot y = ry$,
with $r$ an irrational number.
Its integral curves are dense immersions $\R \to \mathbb{T}^2$.
These immersed submanifolds constitute a foliation
of the torus invariant by the flow.
However,
there is no diffeomorphism $\R \times N \to \mathbb{T}^2$,
as well as no nontrivial constants of the motion.
\medskip

In the usual Hamilton--Jacobi theory
a family of vector fields is usually determined
by solving an associated partial differential equation
of first order.
This requires the use of a skew-symmetric $(0,2)$-tensor field
which relates a vector field, say $Z$, with a 1-form.
The skew-symmetry ensures that
the contraction of $Z$ with the corresponding 1-form
identically vanishes.

\section{Slicing of Hamiltonian systems}

In the standard Hamilton--Jacobi theory
the skew-symmetric $(0,2)$-tensor is assumed to be
the natural symplectic structure of the cotangent bundle.
The classical Hamilton--Jacobi theory makes an essential use
of a symplectic structure.
In view of this,
we still consider the most general slicing problem
but now for  a Hamiltonian system.
Thus $P$ is endowed with a symplectic form~$\omega$,
which defines a vector bundle isomorphism
$\widehat \omega \colon \Tan P \to \Tan^*P$;
and $Z = Z_H$ is the Hamiltonian vector field
of a Hamiltonian function $H \colon P \to \R$:
$Z = \widehat\omega^{-1} \comp \d H$.

\begin{lemma}
Consider a Hamiltonian dynamical system
$(P,\omega,H)$
and $Z = Z_H$ its Hamiltonian dynamical vector field.
Let $\alpha \colon M \to P$ be a map,
and $X$ an arbitrary vector field on~$M$.
We have the following relations:
$$
{}^{t}(\Tan \alpha)
\comp
\widehat\omega
\comp
\Tan \alpha \comp X
=
i_X \alpha^*(\omega)
\,,
$$
$$
{}^{t}(\Tan \alpha)
\comp
\widehat\omega
\comp
Z \comp \alpha
=
\d \,\alpha^*(H)
\,,
$$
where all the vector bundle sections and maps
are understood to be over the base space~$M$.
\end{lemma}
These relations are expressed in the following diagram
(we insist that, since we have to work with the transpose morphism
${}^{t}(\Tan \alpha)$,
all the involved vector bundles are considered
over the base space~$M$):
$$
\xymatrix{
*++{\Tan M}  \ar[d]_{}   \ar[r]^{\Tan \alpha \quad}
&
*++{M \times_\alpha \Tan P}  \ar[r]^{\widehat\omega}
&
*++{M \times_\alpha \Tan^*P} \ar[r]^{\quad {}^{t}(\Tan \alpha)}
&
*++{\Tan^*M}
\\
*++{M}  
\ar[ru]^{\Tan \alpha \comp X \;}_{\; Z \comp \alpha}
\ar@/_2mm/[urrr]^{ i_X \alpha\!^*(\omega) \quad}_{\qquad \d \,\alpha\!^*(H)}
}
$$

\proof
The map
$
\Tan \alpha \comp X
$
is a vector field along~$\alpha$,
$\widehat\omega \comp \Tan \alpha \comp X$
is a differential 1-form along $\alpha$,
and finally
its composition with the transpose morphism
${}^{t}(\Tan \alpha)$
(along~$M$),
$
{}^{t}(\Tan \alpha)
\comp
\widehat\omega
\comp
\Tan \alpha \comp X
$,
is the differential 1-form on~$M$
$
i_X \alpha\!^*(\omega)
$,
since
$
{}^{t}(\Tan \alpha)
\comp
\widehat\omega
\comp
\Tan \alpha
=
\widehat{\alpha^*(\omega)}
$.

On the other hand,
since $Z$ is the Hamiltonian vector field of~$H$,
$
\widehat\omega \comp Z \comp \alpha
=
\d H \comp \alpha
$,
a differential 1-form along $\alpha$,
and its composition with the transpose morphism
${}^{t}(\Tan \alpha)$
is just de pull-back by~$\alpha$ of $\d H$,
$
{}^{t}(\Tan \alpha)
\comp
\widehat\omega
\comp
Z \comp \alpha
=
\alpha\!^*(\d H)
=
\d \,\alpha\!^*(H)
$.
\qed

\begin{prop}
\label{prop-ham}
With the preceding notations, if
$(M,\alpha,X)$ is a solution of the slicing equation
for $(P,Z)$,
$
\Tan \alpha \comp X - Z \comp \alpha = 0
$,
then
\beq
i_X \alpha^*(\omega) - \d \,\alpha^*(H) = 0
\,.
\label{basicham}
\eeq
\end{prop}
\proof
From the preceding lemma we have
\beq
\label{mainrel}
{}^{t}(\Tan \alpha)
\comp
\widehat\omega
\comp
(\Tan \alpha \comp X - Z \comp \alpha)
=
i_X \alpha^*(\omega) - \d \,\alpha^*(H)
\,.
\vadjust{\kern -3ex}
\eeq
\qed

Notice by the way that,
if $\alpha\!^*(\omega)$ were a symplectic form on~$M$,
then equation (\ref{basicham}) would mean that
$X$ is the Hamiltonian vector field associated with
the Hamiltonian function $\alpha\!^*(H)$.

\paragraph{Coordinate expressions}
It is interesting to reproduce the proof of the previous equations
in coordinates.
Again we have local charts
$(x^i)$ in~$M$,
$(z^k)$ in~$P$,
and use them to express
$\alpha(x) = (a^k(x))$
and
$X = X^i \,\tanvec{x^i}$.
The symplectic form reads
$\ds
\omega =
\frac12 \,\omega_{k\ell} \, \d z^k \wedge \d z^\ell
$,
where
$\Omega = (\omega_{k\ell})$
is skew-symmetric.
The matrix of $\widehat\omega$ is $\Omega^\top$.
And the Hamiltonian vector field
$\ds
Z = Z_H =
\derpar{H}{z^\ell} \, \omega^{\ell k} \tanvec{z^k}
$,
where $(\omega^{k\ell}) = \Omega^{-1}$.

Then
$
\Tan \alpha \comp X - Z \comp \alpha
$
in coordinates reads
$$
(x^i) \mapsto
\left(
a^k(x) ;
\derpar{a^k}{x^i} X^i -
\derpar{H}{z^\ell}(\alpha(x)) \, \omega^{\ell k}(\alpha(x))
\right)
\,.
$$
And
$\ds
\alpha^*(\omega) =
\frac12 \, \omega_{k\ell}(\alpha(x)) \,
\derpar{a^k}{x^i} \, \derpar{a^\ell}{x^j} \,
\d x^i \! \wedge \d x^j
$,
$\ds
i_X \alpha^*(\omega) =
X^i \,
\omega_{k\ell}(\alpha(x)) \,
\derpar{a^k}{x^i} \derpar{a^\ell}{x^j} \,
\d x^j
$,
$\ds
\d \alpha^*(H) =
\derpar{H}{z^k}(\alpha(x)) \, \derpar{a^k}{x^j} \, \d x^j
$,
so that
$
i_X \alpha^*(\omega) - \d \,\alpha^*(H)
$
reads
$$
\left(
X^i \,
\omega_{k\ell}(\alpha(x)) \,
\derpar{a^k}{x^i} \derpar{a^\ell}{x^j}
 -
\derpar{H}{z^k}(\alpha(x)) \derpar{a^k}{x^j}
\right)
\d x^j
\,.
$$
We see that multiplying the local expression of
$
\Tan \alpha \comp X - Z \comp \alpha
$
by
$(\omega_{k\ell})$
and then by
$(\derpar{a^\ell}{x^j})$
we obtain the local expression of
$
i_X \alpha^*(\omega) - \d \,\alpha^*(H)
$.

\medskip
In general the morphism ${}^{t}(\Tan \alpha)$ is not bijective,
therefore the implication in the previous proposition
cannot be inverted.
This is easily seen in an example.

\paragraph{Example}
Consider a Hamiltonian system $(P,\omega,H)$.
Let $\alpha \colon I \to P$ be any path that
is \emph{not} a solution of the Hamilton's equation,
but such that $H \comp \alpha = \mathrm{const}$,
and consider the vector field $X = {\d \over \d t}$
on $I \subset \R$.
Then
$
i_X \alpha^*(\omega) - \d \,\alpha^*(H)
$
vanishes trivially,
whereas, of course,
$
\Tan \alpha \comp X - Z \comp \alpha =
\alpha' - Z \comp \alpha \neq 0
$.
\medskip

The preceding proposition gives a link between
the slicing problem
and the usual formulation of the
Hamilton--Jacobi equation, $\d\,\alpha^*(H) = 0$.
However, we need to revert the direct implication,
and this can be done in some cases,
as it was already shown in our paper~\cite{HJ}.
There are at least two ways for doing this,
according to whether we have an isotropy condition, as below,
or a fibred structure, as in the next section.

\subsection*{Isotropic and Lagrangian embeddings}

In this subsection we are going to study slicings
satisfying a geometric property with respect to the
symplectic form.
First we need to recall that
a submanifold $M \subset P$ is called
isotropic, coisotropic or Lagrangian \cite{We73}
when all the tangent spaces at each point are,
which means:
\begin{itemize}
\itemsep 0pt
\item
isotropic:
$\Tan_zM \subset (\Tan_zM)^\bot$;
\item
coisotropic:
$(\Tan_zM)^\bot \subset \Tan_zM$;
\item
Lagrangian:
isotropic and coisotropic:
$\Tan_zM = (\Tan_zM)^\bot$.
\end{itemize}
(Here the orthogonality is taken with respect to the symplectic form.)

An important type of solutions $\alpha \colon M \to P$
of the slicing equation for~$Z$ satisfy
the condition
\beq
\alpha^*(\omega) = 0
\,.
\label{isotropy}
\eeq
When $\alpha$ has constant rank
this condition means that, locally, the image
$\alpha(M) \subset P$
is an \emph{isotropic submanifold}.
When $\alpha$ is an immersion this requires that
$\dim M \leq \frac12 \dim P$.
Of course, in this case the preceding proposition takes a simpler
form:
a solution of the slicing problem
satisfies
$
\d \,\alpha^*(H) = 0
$,
whereas its converse is false,
as is also shown by the same preceding example.

To go further, we need a couple of lemmas.
The notation $F^\circ \subset E^*$ denotes the \emph{annihilator}
of a vector subspace $F \subset E$.

\begin{lemma}
Suppose that $\alpha$ is an embedding,
so that $P_0 = \alpha(M) \subset P$ is a submanifold.
Then:
\begin{itemize}
\itemsep 0pt
\item
$\alpha^*(\omega) = 0$
iff
$\widehat\omega(\Tan P_0) \subset (\Tan P_0)^\circ$,
\emph{i.e.}, $P_0 \subset P$ is an isotropic submanifold.
\item
$\widehat\omega(\Tan P_0) = (\Tan P_0)^\circ$
iff
$\widehat\omega(\Tan P_0) \subset (\Tan P_0)^\circ$
and $\dim P = 2 \dim M$,
\emph{i.e.}, $P_0 \subset P$ is a Lagrangian submanifold.
\end{itemize}
\end{lemma}
\proof
The first statement is a consequence of the fact that
$\alpha^*(\omega)$
is essentially the restriction of $\omega$
to tangent vectors to $\alpha(M)$;
the second one is a matter of dimension counting: $m = p-m$.
\qed

\begin{lemma}
If $\alpha$ is an embedding with $\alpha(M) = P_0$ then
$\Ker {}^{t}(\Tan \alpha) = (\Tan P_0)^\circ$.
\end{lemma}
\proof
Basic linear algebra applied to $\Tan_x\alpha$ for every $x \in M$.
\qed

Consider the following diagram, which contains all of these objects:
$$
\xymatrix{
*++{M \!\times_\alpha\! \Tan P_0}
\ar@{^{(}->}[d]
&
*++{(M \!\times_\alpha\! \Tan P_0)^\circ}
\ar@{^{(}->}[d]
\\
*++{M \times_\alpha \Tan P}  \ar[r]^{\widehat\omega}
&
*++{M \times_\alpha \Tan^*P} \ar[r]^{\quad {}^{t}(\Tan \alpha)}
&
*++{\Tan^*M}
\\
*++{M} 
\ar[u]_{Z \comp \alpha}
\ar[ur]_{\d H \comp \alpha}
\ar@/_3mm/[urr]_{\qquad \alpha\!^*(\d H) \,=\, \d \,\alpha\!^*(H)}
}
$$

\begin{teor}
Let $(P,\omega,H)$ be a symplectic Hamiltonian system,
with Hamiltonian vector field~$Z$,
and let $\alpha \colon M \to P$ be an embedding.
\\
If $\alpha$ is a solution of the slicing equation~(\ref{slicing})
(that is, $Z$ is tangent to $\alpha(M)$)
and satisfies the isotropy condition
($\alpha^*(\omega) = 0$)
then $\alpha$ satisfies
\beq
\d \,\alpha^*(H) = 0
\,.
\label{lagranslicing}
\eeq
Conversely, if $\alpha$ satisfies this equation
and the Lagrangianity condition
($\alpha^*(\omega) = 0$ and $\dim P = 2 \dim M$)
then it is a solution of the slicing equation.
\end{teor}
\proof
We have already proved the direct implication.

Conversely, if $\alpha\!^*(\d H)$ is zero then
$\d H \comp \alpha$ takes its values in the kernel,
which is
$(M \!\times_\alpha\! \Tan P_0)^\circ$.
When the Lagrangianity condition
$\widehat\omega(\Tan P_0) = (\Tan P_0)^\circ$
holds we conclude that
$Z \comp \alpha$ is a section of $\Tan P_0$,
or, in other words,
that $Z$ is tangent to $P_0$,
which is one of the ways of saying that $\alpha$ is slicing of~$Z$.
\qed

So for Lagrangian embeddings
to solve the slicing equation
is equivalent to solving equation~(\ref{lagranslicing}).
We call these solutions \emph{Lagrangian slicings} of~$Z$.

\subsection*{Constants of the motion and involutivity}

In the preceding section we have observed
the close relationship between complete slicings
and constants of the motion.
So, consider a submersion $F \colon P \to \R^n$,
with level sets $P_c \equiv F^{-1}(c)$.

\begin{lemma}
The functions $F^i$ are in involution,
$\{F^i,F^j\}=0$,
iff
all the level sets $P_c$
are coisotropic submanifolds of~$P$.

When $\dim P = 2n$ this means that the $P_c$ are
Lagrangian submanifolds.
\end{lemma}
\proof
The proof is easy, see
\cite[p.\,101]{LM}.
\qed

So,
a complete slicing given by
$n$ constants of the motion \emph{in involution}
has coisotropic leaves,
and if $\dim P = 2n$
then the leaves are Lagrangian submanifolds,
and conversely.

\paragraph{Remark}
As all our preliminary analysis has been made without
the help of a $(0,2)$-tensor field,
it is clear that when the vector field $Z$ allows
for alternative invariant skew symmetric $(0,2)$-tensor fields,
it is possible to consider alternative cotangent bundle structures
on $P$ and therefore different projections.

\paragraph{Example} 
We can consider the isotropic harmonic oscllator and if we write, 
in coordinates 
$(x,p) \in \R^2$,
$$
x\,\cos \alpha + P\, \sin\alpha = q
\,,  \quad 
P\,\cos \alpha -x\, \sin\alpha=p
\,,
$$
we have that 
$\d q \wedge \d p = \d x \wedge \d P$, 
$\d (p\,\d q) = \d(P\,\d x)$.

The fibering vector fields 
$p \,\partial/\partial p$ and 
$P \,\partial/\partial P$ 
are diffeomorphically related but induce 
alternative cotangent bundle structures on $\mathbb{R}^2$
\cite{CIMM}


\subsection*{Local existence of complete Lagrangian slicings}

In the preceding section we have proved a
local existence theorem for complete slicings.
Now we are going to prove a similar result
in the Hamiltonian framework,
for solutions satisfying the Lagrangianity condition.
See also
\cite[p.\,156]{Sch}.

\begin{teor}
Let $(P,\omega,H)$ be a symplectic Hamiltonian system,
with Hamiltonian vector field~$Z$,
and $z_\cc \in P$ a \emph{noncritical} point of~$H$.
There exists a Lagrangian slicing of~$Z$
passing through~$z_\cc$.
Indeed, this slicing is contained in
a local complete Lagrangian slicing of~$Z$.
\end{teor}

\proof
By applying the Carath\'eodory--Jacobi--Lie theorem
---see for instance
\cite[p.\,51]{LM}---
$H$ can be included in a set of local Darboux coordinates
$(q^1,\ldots,q^n,H=p_1,\ldots,p_n)$
centered at~$z_\cc$.
Then $Z = \derpar{}{q^1}$
is tangent to the Lagrangian submanifolds of~$P$ defined by
$p_1 = c_1$, \ldots\ , $p_n = c_n$.
These submanifolds constitute the complete slicing we sought.
\qed

\subsection*{Poisson Hamiltonian systems}

The preceding argument can be adapted to the Poisson case.
Let $P$ be a manifold endowed with an almost-Poisson tensor field
$\Lambda$, that is to say,
a section of $\mathsf{\Lambda}^2 \Tan P$.
This defines a vector bundle morphism
$
\widehat\Lambda \colon \Tan^*P \to \Tan P
$
by
$
\left\langle \beta , \widehat\Lambda(\alpha) \right\rangle
=
\Lambda(\alpha,\beta)
$.
The image of this morphism,
$C = \mathrm{Im} \widehat\Lambda \subset \Tan P$,
is called the
\emph{characteristic tangent distribution}
of~$\Lambda$.
If $\Lambda$ has constant rank then $C$ is a vector subbundle.

We will need a generalisation of the concept of
Lagrangian submanifold to the Poisson case.
A submanifold $P_0 \subset P$ of an almost-Poisson manifold
is called \emph{Lagrangian}
\cite[p.\,100]{Vai}
when
$$
\widehat\Lambda
\left(
  (P_0 \times_{P_0} \Tan P_0)^\circ
\right)
=
\Tan P_0 \cap (P_0 \times_{P_0} C)
\,.
$$

The almost-Poisson tensor field also defines an almost-Poisson bracket
$
\{f,g\} = \Lambda (\d f ,\d g)
$,
which is skew-symmetric
and a derivation on each of its arguments
(it does not necessarily satisfy the Jacobi identity
unless the Schouten bracket vanishes, i.e., 
$[\Lambda,\Lambda]=0$).

Suppose that we have a Hamiltonian function
$H \colon P \to \R$,
which defines a Hamiltonian vector field
$Z = Z_H = \widehat\Lambda \comp \d H$
and the corresponding Hamiltonian dynamics.
We want to study the slicing problem for $(P,Z)$.

As before, we consider the elements in this diagram,
but notice that $\Lambda$ may be degenerate:
$$
\xymatrix{
*++{M \!\times_\alpha\! \Tan P_0}
\ar@{^{(}->}[d]
&
*++{(M \!\times_\alpha\! \Tan P_0)^\circ}
\ar@{^{(}->}[d]
\\
*++{M \!\times_\alpha\! \Tan P}
&
*++{M \!\times_\alpha\! \Tan^*P}
\ar[l]_{\widehat\Lambda}
\ar[r]^{\qquad {}^t(\Tan \alpha) \quad}
&
*++{\Tan^*M}
\\
&
*++{M}
\ar@/^2mm/[ul]^{Z \comp \alpha}
\ar[u]_{\d H \comp \alpha}
\ar@/_2mm/[ur]_{\quad \alpha\!^*(\d H) \,=\, \d \,\alpha\!^*(H)}
}
$$

\begin{lemma}
Let $E$ be a finite-dimensional vector space,
$E^*$ its dual space,
$E_0 \subset E$ a vector subspace,
$\lambda \colon E^* \to E$ a linear map,
$\delta \in E^*$ a covector.
Denote by
$E_0^\circ \subset E^*$
the annihilator of~$E_0$
and by
${}^t\lambda \colon E^* \to E$
the transpose map of~$\lambda$.
Then
$\lambda(\delta) \in E_0$
iff
$\delta \in ({}^t\lambda(E_0^\circ))^\circ$.
If moreover $\lambda$ is
symmetric or skew-symmetric
(${}^t \lambda = \pm\lambda$),
then
$\lambda(\delta) \in E_0$
iff
$\delta \in (\lambda(E_0^\circ))^\circ$.
\qed
\end{lemma}

\begin{teor}
Let $(P,\Lambda,H)$ be an almost-Poisson Hamiltonian system,
with Hamiltonian vector field~$Z$.
Let $\alpha \colon M \to P$ be an embedding
with image
$\alpha(M) = P_0$.
Then $\alpha$ is a solution of the slicing equation
iff
\beq
\d H \comp \alpha
\hbox{ is a section of }
\left(
\widehat\Lambda\left(
  (M \times_\alpha \Tan P_0)^\circ
\right)
\right)^\circ
.
\eeq
Suppose that $P_0 \subset P$ is a \emph{Lagrangian} submanifold.
Then $\alpha$ is a solution of the slicing equation
iff
\beq
\d H \comp \alpha
\hbox{ is a section of }
(M \times_\alpha \Tan P_0)^\circ + (M \times_\alpha \Ker\widehat \Lambda)
,
\eeq
that is to say,
iff
\beq
\alpha\!^*(\d H)
\hbox{ is a section of }
{}^t(\Tan \alpha) (\Ker\widehat \Lambda)
.
\eeq
\end{teor}
\proof
The first statement is a consequence of the lemma.

As for the second statement,
being $P_0$ Lagrangian means that
$
\widehat\Lambda\left(
  (P_0 \times_{P_0} \Tan P_0)^\circ
\right)
=
\Tan P_0 \cap (P_0 \times_{P_0} C)
$.
When restricted this to $\alpha$ and with the annihilator we have
$
\left(
\widehat\Lambda\left(
  (M \times_\alpha \Tan P_0)^\circ
\right)
\right)^\circ
=
\left(
  (M \times_\alpha \Tan P_0) \cap (M \times_\alpha C)
\right)^\circ
=
  (M \times_\alpha \Tan P_0)^\circ + (M \times_\alpha C)^\circ
$,
and remember that
$
C^\circ = \Ker {}^t\widehat \Lambda = \Ker \widehat \Lambda
$.
\qed

The symplectic case is obtained when $C = \Tan P$,
or equivalently when $\Ker \widehat \Lambda = \{0\}$.
Then for the Lagrangian case
the last statement in the theorem means that
$\alpha$ is a slicing iff
$\alpha\!^*(\d H) = 0$,
as was already given by theorem~3.

\paragraph{Example}
Consider
$P = \R^3$
with coordinates $(x,y,z)$
and the Poisson structure given by the Poisson bracket
$\ds
\{f,g\} =
z \left(
  \derpar{f}{y} \derpar{g}{x} - \derpar{f}{x} \derpar{g}{y}
\right)
$
---indeed,
this is the Lie--Poisson structure constructed
from the Heisenberg Lie algebra
\cite[p.\,153]{Vai}.
The Hamiltonian function
$
H = \frac12 z(x^2+y^2)
$
defines the Hamiltonian vector field
$\ds
Z =
z^2 \left( -y \derpar{}{x} + x \derpar{}{y} \right)
$.

We have two constants of the motion,
${x^2+y^2}$ and~$z$.
Excluding the $z$-axis,
all their level sets are diffeomorphic to the unit circle;
parametrising the circle with the natural angle,
these diffeomorphisms read
$\alpha_{r,c}(\phi) = (r \cos\phi, r\sin\phi, c)$.
It is easily checked that
$\alpha_{r,c}^*(\d H) = 0$.

Since for $c \neq 0$
all these level sets are Lagrangian submanifolds,
we conclude from the preceding theorem
that the $\alpha_{r,c}$ constitute
a complete Lagrangian slicing for~$Z$
on the open set given by $z \neq 0$.

\medskip

In general this situation prevails for Poisson manifolds
and we have to consider Casimir functions and constants of
the motion in involution.
Casimir functions identify `parameters'
(like mass, spin, charge, isospin, coloured charge)
while the constants of the motion identify the decomposition
into vector fields on lower dimensional submanifolds

\section{Slicing in fibred manifolds}

In this section we consider a dynamical system $(P,Z)$,
where the manifold $P$ is fibred over another manifold,
that is to say,
we work in a
\emph{fibre bundle}
$\pi \colon P \to M$.
We consider the slicing problem as before:
$$
\xymatrix{
*++{\Tan M}  \ar[r]^{\Tan \alpha}  \ar[d]_{} &
*++{\Tan P}  \ar[d]_{}
\\
*++{M} \ar[r]^{\alpha} \ar@/^3mm/[u]^{X} &
*++{P} \ar@/^3mm/[l]^{\pi} \ar@/_3mm/[u]_{Z}
}
$$
but only for \emph{sections} of~$\pi$,
that is to say,
for maps
$\alpha \colon M \to P$
such that $\pi \comp \alpha = \mathrm{Id}_M$.
For this problem there is not a `gauge freedom' as mentioned
in section~2:
the submanifold $\alpha(M) \subset P$
cannot be expressed as the image of any other section.

Since $\alpha$ is an embedding,
we know that equation (\ref{slicing}) determines~$X$.
Nevertheless, composing this equation with the tangent map
$\Tan \pi$,
we can give an explicit formula for~$X$:
\beq
X = \Tan \pi \comp Z \comp \alpha
\,.
\label{Xfibred}
\eeq
So, from now on we assume that $X$ is defined by this equation
from~$\alpha$.
In this case, proposition~1 adopts the following form:
\begin{prop}
A section $\alpha$ of $\pi \colon P \to M$
is a solution of the slicing equation for $(P,Z)$
iff
\beq
\Tan \alpha \comp \Tan \pi \comp Z \comp \alpha = Z \comp \alpha
\,;
\label{basicfib}
\eeq
that is to say,
if $\Tan \alpha \comp \Tan \pi \comp Z$ agrees with~$Z$
on the submanifold $\alpha(M)$.
\qed
\end{prop}

\begin{lemma}
If $\alpha$ is a slicing section,
the vector field along~$\alpha$
defined as
$\Tan \alpha \comp \Tan \pi \comp Z \comp \alpha - Z \comp \alpha$
is $\pi$--vertical.
\end{lemma}
Remember that the vertical subbundle of $\Tan P$
is
$
\mathrm{V} P = \Ker \Tan \pi
$.
Its fibres are
$
\mathrm{V}_z P =
\Ker \Tan_z \pi \subset
\Tan_zP$
and are naturally identified with the tangent spaces to the fibres
of~$\pi$.
Application of $\Tan \pi$ to
$\Tan \alpha \comp \Tan \pi \comp Z \comp \alpha - Z \comp \alpha$
yields immediately zero since $\alpha$ is a section of~$\pi$.
\qed

\subsection*{Sections, projectors, and connections}

If $\alpha$ is a section of~$P$,
let us have a look at the composition
$\Tan \alpha \comp \Tan \pi$.
At a given point $z = \alpha(x) \in P$,
$\Tan_z(\alpha \comp \pi) \colon \Tan_zP \to \Tan_zP$
is an endomorphism,
and since $\Tan \pi \comp \Tan \alpha$ is the identity,
we note that
$
\Tan_z(\alpha \comp \pi) \comp \Tan_z(\alpha \comp \pi) =
\Tan_z(\alpha \comp \pi)
$,
therefore it is a projector in $\Tan_zP$.
Since $\Tan_x\alpha$ is injective, it is clear that
$$
\Ker \Tan_z(\alpha \comp \pi) =
\Ker \Tan_z \pi =
\mathrm{V}_z P
\,.
$$
Therefore
$$
\mathrm{Im} \Tan_z(\alpha \comp \pi) =
\Tan_{z}\alpha(M)
$$
is a complementary subspace to $\mathrm{V}_z P$.

So we can write, for every $x \in M$, a direct sum decomposition
$$
\Tan_{\alpha(x)} P =
\Ver_{\alpha(x)}P \,\oplus\, \Tan_{\alpha(x)} \alpha(M)
\,.
$$
This can be written globally in the pull-back vector bundle:
$$
M \!\times_\alpha\! \Tan P =
M \!\times_\alpha\! \Ver P
\,\oplus\,
M \!\times_\alpha\! \Tan \,\alpha(M)
\,.
$$

Now suppose that we have not only a section
but a family of non overlapping sections covering the whole
manifold~$P$;
this can be defined by a diffeomorphism
$\overline\alpha \colon M \times N \to P$,
where each $\alpha_c = \overline\alpha(\cdot,c)$ is a section of~$P$,
but this diffeomorphism could as well be defined on open sets of~$P$.
The preceding study can be performed at every point $z \in P$,
therefore the family $\overline\alpha$ defines a
\emph{horizontal subbundle},
that is,
a vector subbundle $H \subset \Tan P$
complementary to the vertical subbundle
$\Ver P \subset \Tan P$.
A horizontal subbundle of $\Tan P$ is also called
a (nonlinear) \emph{connection} on the bundle~$P$.
This horizontal subbundle is obviously integrable,
its integral manifolds being given by the embeddings $\alpha_c$.

Conversely,
if a connection on the bundle
$P \to M$
has integrable horizontal subbundle
(which amounts to saying that its curvature vanishes,
see
\cite[p.\,90]{Sau}),
then
its integral manifolds are locally the images of sections of
the bundle.

\subsection*{Complete solutions and connections}

Still working with the diffeomorphism
$\overline\alpha \colon M \times N \to P$,
when is it a complete solution of the
slicing equation for sections?
In addition to defining an integrable horizontal subbundle,
$Z$ has to be tangent to it.
Therefore, locally,
complete solutions of the slicing equation
are equivalent to
connections on $\pi \colon P \to M$,
with zero curvature,
and invariant by~$Z$.

\subsection*{The Hamiltonian case on a fibred manifold}

Here we consider both a bundle structure and a Hamiltonian structure
on~$P$.
So, $\pi \colon P \to M$ is a fibre bundle
and $(P,\omega)$ is a symplectic manifold,
and $Z = Z_H$ is a Hamiltonian vector field
(with Hamiltonian function~$H$).
Let $\alpha$ be a section of~$\pi$,
and let us determine if it is a slicing section for~$Z$.
We wish to give a kind of converse to
proposition~\ref{prop-ham},
which relates
$
\Tan \alpha \comp X - Z \comp \alpha
$
with
$
i_X \alpha\!^*(\omega) - \d \,\alpha\!^*(H)
$
(where $X$ is given by
$X = \Tan \pi \comp Z \comp \alpha$.)
$$
\xymatrix{
*++{\Tan M}  \ar[d]_{}   \ar[r]^{\Tan \alpha \quad}
&
*++{M \!\times_\alpha\! \Tan P}  \ar[r]^{\widehat\omega}
&
*++{M \!\times_\alpha\! \Tan^*P} \ar[r]^{\quad {}^{t}(\Tan \alpha)}
&
*++{\Tan^*M}
\\
*++{M} 
\ar[ru]_{\; \Tan \alpha \comp X - Z \comp \alpha}
\ar@/_3mm/[urrr]_{\qquad i_X \alpha\!^*(\omega) - \d \,\alpha\!^*(H)}
}
$$
In this diagram
$\widehat \omega$ is bijective,
and, as we have already noted, the problem is that
${}^t(\Tan \alpha)$
is not injective,
since
$\Ker {}^t(\Tan \alpha) = (M \!\times_\alpha\! \Tan P_0)^\circ$.
However, we have also noted that
$\Tan \alpha \comp X - Z \comp \alpha$
is $\pi$--vertical.
Therefore we only need to impose the injectivity of
the restriction of
${}^t(\Tan \alpha) \comp \widehat\omega$
to the subbundle
$M \!\times_\alpha\! \Ver P$,
and this is equivalent to saying that
$$
\widehat\omega ( M \!\times_\alpha\! \Ver P )
\cap
(M \!\times_\alpha\! \Tan P_0)^\circ
=
\{0\}
\,.
$$

\begin{lemma}
With the preceding hypotheses,
the following conditions are equivalent:
\bit
\itemsep 0pt plus 1pt
\item
The fibres of $\pi \colon P \to M$
are isotropic submanifolds (with respect to~$\omega$).
\item
For every couple of vertical vectors
$w_z,w'_z \in \Ver_zP \subset \Tan_zP$
one has
$\omega(w_z,w'_z) = 0$.
\item
$\widehat\omega(\Ver P) \subset (\Ver P)^\circ$.
\eit
\end{lemma}
\proof
The equivalence of the first two is due to the fact that
the vertical vectors are those that are tangent to the fibres.
\qed

\begin{cor}
If $\alpha$ is a section of~$P$
and the fibres are isotropic
then
$
\widehat\omega ( M \!\times_\alpha\! \Ver P )
\cap
(M \!\times_\alpha\! \Tan P_0)^\circ
=
\{0\}
$.
\end{cor}
\proof
The vertical+horizontal decomposition yields
$
M \!\times_\alpha\! \Tan^*P =
(M \!\times_\alpha\! \Ver P)^\circ \oplus
(M \!\times_\alpha\! P_0)^\circ
$.
\qed

\begin{teor}
\label{teor-fibham}
Let $(P,\omega,H)$ be a Hamiltonian system
on a fibre bundle $\pi \colon P \to M$.
Let $\alpha \colon M \to P$
be a section of~$\pi$,
and define its associated vector field
$X = \Tan \pi \comp Z \comp \alpha$.
Suppose that the fibres of~$\pi$ are isotropic.
Then $\alpha$ is a slicing section
iff
$$
i_X \alpha^*(\omega) - \d \,\alpha^*(H) = 0
\,.
$$
\end{teor}
\proof
As we have just shown, the isotropy condition
implies that
${}^t(\Tan \alpha) \comp \widehat\omega$
is injective when applied to vertical vectors.
Therefore
if
$
i_X \alpha^*(\omega) - \d \,\alpha^*(H)
$
is zero
then
$\Tan \alpha \comp X - Z \comp \alpha$
also is.
\qed

\paragraph{Coordinate expressions}
Let's understand the proof of the theorem
on the light of coordinates.
We use coordinates
$(x^i)$ in~$M$
and adapted coordinates
$(x^i,y^\mu)$ in~$P$.
The section takes the form
$\alpha(x) = (x,a^\mu(x))$
and its tangent map is represented by the matrix
$
\left(
I \atop A
\right)
$,
where $A$ is the jacobian matrix of the $a^\mu$.
The symplectic form $\omega$ is represented by a skew-symmetric matrix
$
\Omega =
\small
\left( \begin{array}{cc}
\Omega_b & N
\\
-N^\top & \Omega_f
\end{array} \right)
$.
The matrix of $\widehat\omega$ is $\Omega^\top$.
Then the linear map
$
{}^t(\Tan_x\alpha) \comp \widehat \omega_z
$
is represented by the matrix
$\small
\left( \begin{array}{cc}
\Omega_b^\top + A^\top N^\top
&
-N + A^\top \Omega_f^\top
\end{array} \right)
$,
and its restriction to the vertical subspace by its second block,
$$
-N + A^\top \Omega_f^\top
\,.
$$
Now,
the fibres are isotropic iff $\Omega_f = 0$,
and since $\Omega$ is nondegenerate
$N$ has to have maximal rank and be injective.
So, the only vertical vector sent to 0 by this map is 0.
\medskip

In the preceding section we have already obtained
the equation for the Lagrangian slicings.
We can combine theorems 3 and~6 in this way:
\begin{cor}
For a Hamiltonian system $(P,\omega,H)$ fibred over~$M$,
with isotropic fibres,
let $\alpha \colon M \to P$ be a section with
isotropic image.
Then $\alpha$ is a solution of the slicing problem
iff
$$
\d \,\alpha^*(H) = 0
\,.
$$
\end{cor}
\proof
The isotropy of the fibres requires
$\dim M \geq \dim P \,/\,2$
and the isotropy of $\alpha(M)$ requires
$\dim M \leq \dim P \,/\,2$.
Therefore
$\dim M = \dim P \,/\,2$,
which in particular means that $\alpha(M) \subset P$
is a Lagrangian submanifold
and then application of theorem~2 yields the desired result.
Otherwise,
the isotropy of the image means
$\alpha^*(\omega) = 0$
and one can apply theorem~4 at once.
\qed

The isotropy of the fibres is necessary to prove this result,
as shown by the following example.
\paragraph{Example}
Take
$P = \R^4$,
with coordinates
$(x,p_x,y,p_y)$,
with the usual symplectic form
$\omega = \d x \wedge \d p_x + \d y \wedge \d p_y$,
and the Hamiltonian of the isotropic double harmonic oscillator
$H = \frac12( x^2 + p_x^2 + y^2 + p_y^2)$;
its Hamiltonian vector field is
$Z =
p_x \,\tanvec{x} - x \,\tanvec{p_x} + p_y \,\tanvec{y} - y \,\tanvec{p_y}
$.

Consider the trivial fibre bundle
$\pi \colon P \to M$
given by
$M = \R$,
with projection
$\pi(x,p_x,y,p_y) = x$.
Of course, since $M$ is 1-dimensional
any section $\alpha$ of~$\pi$ satisfies $\alpha^*(\omega)=0$.

Then consider the local section
$\alpha(x) = (x,x,\sqrt{c^2-x^2},\sqrt{c^2-x^2})$.
It satisfies
$H \comp \alpha = c^2 = \mathrm{const}$,
but one easily checks that it is not a slicing section.
The point is that the fibres of~$\pi$ are not isotropic
---they cannot be since they are 3-dimensional submanifolds
of a 4-dimensional symplectic manifold.

\section{Lagrangian and Hamiltonian formalisms}

In this section we study some features specific to the dynamics
on tangent and cotangent bundles,
and in particular to Lagrangian and Hamiltonian formalisms.

\medskip

First, notice that
the results of the preceding section 
apply directly to a canonical Hamiltonian system
$(P=\Tan^*Q,\omega,H)$,
whith $\Tan^* Q$ endowed with its 
vector bundle structure 
$\pi \colon \Tan^*Q \to Q$
and its canonical symplectic form~$\omega$.
The dynamical vector field $Z$ is the symplectic gradient
$Z_H$ of the Hamiltonian function~$H$.

Then we consider the slicing equation $X \sim_\alpha Z$
for a section $\alpha$ of~$P$, that is to say,
a differential 1-form on~$Q$.
From 
(\ref{Xfibred}) 
we can compute the slicing vector field~$X$, 
which in this case turns out to be
$X = \FD H \comp \alpha$,
where 
$\FD H \colon \Tan^*Q \to \Tan Q$
is the fibre derivative of~$H$.

Now, notice that the fibres of $\Tan^*Q$,
that is to say, 
the cotangent spaces $\Tan_q^*Q$,
are \emph{isotropic} submanifolds of the cotangent bundle
with respect to its canonical symplectic structure.
So, we are under the hypotheses of 
theorem~\ref{teor-fibham}
and its corollary,
which give a special form for the slicing equation.

In particular, 
the classical \emph{Hamilton--Jacobi equation}
is nothing but
the slicing equation for a \emph{closed} 1-form~$\alpha$.
This means that $\alpha$ is locally exact,
$\alpha = \d W$,
and the slicing equation has the well-known form
\beq
H \comp \d W = \mathrm{const}
\,.
\eeq

\medskip

The same applies to the Lagrangian formulation
of mechanics 
when it is defined by a \emph{regular} Lagrangian function
$L \colon \Tan Q \to \R$.
In this case the fibred manifold is $P = \Tan Q$;
now we don't have a canonical symplectic form,
but the 2-form $\omega_L$ defined from the Lagrangian.
The Hamiltonian vector field is the symplectic gradient
of the energy $E_L$ 
\cite{AM}.
Then all proceeds as in the Hamiltonian case.

\medskip

Within this framework we recover some of our previous results.
In fact 
theorem~\ref{teor-fibham} 
has, as particular cases,
theorems~1 and~2 in our paper 
\cite{HJ},
corresponding to
the Lagrangian and
the Hamiltonian
formulations, respectively.
In the same way, corollary~3 corresponds to
propositions~3 and~7 of the same paper.
There it is also proved (theorem~3) the equivalence between
the Hamilton--Jacobi theories
for the Lagrangian and the Hamiltonian dynamics
for regular systems.
The relationship between constants of the motion and complete slicings
(theorem~\ref{teor-complete-constant})
was also established for these particular cases in~%
\cite{HJ}.

\subsection*{Determination of a second-order dynamics
from constants of the motion}

Suppose we have a foliation 
$\{M_c\}$
of a manifold~$P$.
A vector field 
$Z$ on~$P$ 
tangent to the foliation
defines a vector field
$X_c$ on every leaf $M_c$ of the foliation.
Conversely,
a vector field $X_c$ on every $M_c$ defines a vector field
$Z$ on~$P$
(though a priori one cannot guarantee it to be continuous).

This is what happens when we have a complete slicing 
$\{(M,\alpha_c,X_c)\}$
of a dynamics $(P,Z)$, as discussed in section~3.
Now, suppose that the hypotheses of
theorem~\ref{teor-complete-constant}
are satisfied,
so that the complete slicing is equivalent to a 
(manifold-valued) constant of the motion $F \colon P \to N$.
Then it could seem that 
the dynamics $Z$ is determined by~$F$.
But of course this is not true:
the conditions of the theorem assume that $Z$ is \emph{already} given,
otherwise the vector fields $X_c$ could not be determined.

However, there is a very special instance
where the knowledge of some constants of the motion
suffices to determine the dynamics.
Recall that 
a vector field $Z$ defined on the tangent bundle $\Tan M$
of a manifold 
is said to satisfy the second-order condition when
its integral curves are the velocities of their projections
to the base space~$M$.
It is easily proved that this is equivalent to saying that,
besides being a section of the tangent bundle of $\Tan M$,
$\tau_{\Tan M} \colon \Tan(\Tan M) \to \Tan M$,
$Z$ is also a section of the other vector bundle structure
of $\Tan (\Tan M)$, 
the one given by
$\Tan\tau_{M} \colon \Tan(\Tan M) \to \Tan M$.
In brief, this means that
$\Tan \tau_M \circ Z = \mathrm{Id}$.

\begin{lemma}
Consider a dynamical system
$(P,Z)$
where 
$P \subset \Tan M$ 
is an open subset projecting over~$M$.

If
$(M,\alpha,X)$
is a slicing of~$Z$ by a section $\alpha$ of~$P$,
and $Z$ satisfies the second-order condition, 
then  
$X = \alpha$.

Conversely,
suppose we have a complete slicing 
$(\alpha_c,X_c)$ 
of $Z$ by sections $\alpha_c$ of~$P$.
If $X_c=\alpha_c$ for every~$c$,
then
$Z$ satisfies the second-order condition.
\end{lemma}
\proof
If $Z$ satisfies this condition,
then the slicing equation
$\Tan \alpha \circ X = Z \circ \alpha$,
composed with $\Tan \tau_M$,
yields $X = \alpha$.

Conversely, when $X = \alpha$
the slicing equation reads
$\Tan \alpha \circ \alpha = Z \circ \alpha$,
and composition with $\Tan \tau_M$ yields
$\alpha = \Tan \tau_M \circ Z \circ \alpha$.
This means that $Z$ satisfies the second-order condition 
on every point of $\alpha(M)$.
\qed

\begin{teor}
Let
$P \subset \Tan M$ 
be an open subset projecting onto~$M$.
Suppose we have $m=\dim M$ functions
$f^\alpha \colon P \to \R$
whose fibre derivatives
$\mathcal{F}f^\alpha \colon P \to \Tan^*M$
are linearly independent at each point.

Then around any point $v \in P$ 
there exists a unique local vector field~$Z$,
satisfying the second-order condition,
and for which the $f^\alpha$ are constants of the motion.
\end{teor}
\proof
Put $F = (f^1,\ldots,f^m) \colon P \to \R^m$.
For every $c \in \R^m$ we have a submanifold
$P_c = F^{-1}(c) \subset P$.
The hypotheses imply that the restriction of the projection to 
this submanifold,
$\tau|_{P_c} \colon P_c \to M$,
is a diffeomorphism in a neighbourhood of any point $v \in P_c$.
Let 
$\alpha_c \colon M \to P_c$
be its inverse.
This $\alpha_c$ is also a vector field on~$M$,
so it defines a vector field
$Z|_{P_c}$, 
and this satisfies the second-order condition by the preceding lemma.
All of these together yield~$Z$.

To complete the proof we need to show that $Z$ is smooth,
and we will do this by an explicit computation in coordinates.
Let $v_\circ \in P$ be an arbitrary point,
and use natural coordinates $(q^i,v^i)$ around it.
Write
$\ds
Z = v^i \derpar{}{q^i} + Z^i(q,v) \derpar{}{v^i}
$.
Imposing that the $f^\alpha$ are constants of the motion for~$Z$
we obtain
$$
\Lie_Z f^\alpha = 
\derpar{f^\alpha}{q^i} v^i + 
\derpar{f^\alpha}{v^i} Z^i =
0
\,.
$$
The linear independence of the fibre derivatives means 
in coordinates that the matrix
$\ds
\left(
\derpar{f^\alpha}{v^i}
\right)
$
is invertible.
Hence, we determine the last coefficients of~$Z$ as
$$
Z^i = - 
\bigg(\left(
\derpar{f}{v}
\right)^{-1} 
\bigg)^{\!\!i}_{\beta}
\, \derpar{f^\beta}{q^j}
\, v^j
\,.
\vadjust{\kern -7mm}
$$
\qed

\paragraph{Remark}
It is known that from a vector field $X$ on~$M$
one can construct its canonical lift $X^{\Tan}$
to the tangent bundle. 
This vector field does not satisfy the second-order condition
in the whole $\Tan M$, 
but in the points of $X(M)$ it does.
Indeed, in the first part of the preceding proof,
what we are defining is
$Z|_{X(M)} = X^{\Tan}|_{X(M)}$,
where $\alpha \equiv X$.
Since we have a whole family of $\alpha$'s covering the whole space,
the vector field $Z$ constructed in this way satisfies 
the second-order condition at every point.

\paragraph{Example}
We will use the free particle to show that working in the
Lagrangian or in the Hamiltonian formalisms 
is philosophically different.

If we take 
$P = \Tan^* \R^n$
and the Hamiltonian
$H = \frac12 \left( p_1^2 + \ldots + p_n^2 \right)$
then the dynamical vector field is
$Z = \sum p_i \,\derpar{}{q_i}$,
and its constants of the motion are the functions
$f(p)$.
But notice that these functions are constants of the motion
for $H$ and also for any Hamiltonian of the form
$H(p)$;
the corresponding dynamical vector field is
$Z = \sum \derpar{H}{p_i} \,\derpar{}{q_i}$.

Now take 
$P = \Tan \R^n$
and consider the functions
$f_i = v_i$.
Following the preceding theorem, 
we can look for a second-order vector field $Z$
having the $f_i$ as constants of the motion.
There is a \emph{unique} such a vector field,
and it is
$Z = \sum v_i \,\derpar{}{q_i}$.
By the way, the same vector field would be obtained 
if one considered, instead of the~$v_i$,
any set of $m$ independent functions $f_i(v)$.

\subsection*{Acknowledgments}

JFC and EM 
acknowledge the financial support of 
the Ministerio de Econom\'{\i}a y Competitividad (Spain) project
MTM--2012--33575
and 
the DGA (Aragon) project
DGA E24/1.
XG, MCML and NRR
acknowledge the financial support of the 
Ministerio de Ciencia e Innovaci\'on (Spain) projects 
MTM2014--54855--P and MTM2011--22585.
\bigskip

\def\refname{References}


\end{document}